\documentclass[letterpaper, 10 pt, conference]{ieeeconf}
\IEEEoverridecommandlockouts                              

\usepackage[colorinlistoftodos,obeyFinal]{todonotes}                           
\usepackage{fainekos-macros}
\usepackage{amsmath} 
\usepackage{amssymb}  
\usepackage{graphicx}
\usepackage{tikz}
\usetikzlibrary{arrows}
\usepackage{ifthen}
\usepackage{mathtools}
\usepackage{stmaryrd}
\usepackage{csquotes}
\usepackage{xargs}                      
\usepackage{subcaption}
\usepackage{textcomp}
\usepackage{bm}
\usepackage{cite}
\graphicspath{{./figures/}}
\usepackage{algorithm}
\usepackage[noend]{algpseudocode}
\usepackage{varwidth}

\algrenewcommand\algorithmicforall{\textbf{foreach}}
\algrenewcommand\algorithmicindent{.8em}

\let\oldnl\nl
\newcommand{\nonl}{\renewcommand{\nl}{\let\nl\oldnl}}

\algnewcommand{\LineComment}[1]{\Statex // #1}

\newcommand\restr[2]{{
  \left.\kern-\nulldelimiterspace 
  #1 
  \littletaller 
  \right|_{#2} 
  }}

\usepackage{lipsum}

\usepackage{etoolbox}
\makeatletter
\patchcmd{\maketitle}{\@copyrightspace}{}{}{}
\makeatother

\usepackage{xcolor}

\usepackage{stmaryrd}
\usepackage{hyperref}
\usepackage{bbm}
\newcommand{\ie}{{i.e.,} }
\newcommand{\eg}{{e.g.,} }
\usepackage[automake,toc,abbreviations]{glossaries-extra}
\setabbreviationstyle{long-short}
\glsdisablehyper
\newglossaryentry{computer}
{
name=computer,
description={A programmable machine that receives input data,
               stores and manipulates the data, and provides
               formatted output}
}

\newabbreviation[plural=GPs,longplural=Gaussian Processes]{gp}{GP}{Gaussian Process}

\newabbreviation{mpc}{MPC}{Model Predictive Control}

\newabbreviation{pwr}{PWR}{piecewise residual}

\newabbreviation{mrac}{MRAC}{Model Reference Adaptive Control}

\newabbreviation{motiv_ex}{EX-M}{Motivating example}

\newabbreviation{ddmpc}{DD-MPC}{Data-Driven MPC}

\newabbreviation{nlp}{NLP}{Nonlinear Program}

\newabbreviation{cl}{C.L.}{closed-loop}

\newabbreviation{ol}{O.L.}{open-loop}

\newabbreviation{qp}{QP}{Quadratic Program}

\newabbreviation{lti}{LTI}{linear time-invariant}

\newabbreviation{blr}{BLR}{Bayesian Linear Regression}

\newabbreviation{bnn}{BNN}{Bayesian Neural Network}

\newabbreviation{gmm}{GMM}{Gaussian Mixture Model}

\newabbreviation{cvar}{CVaR}{Conditional Value-at-Risk}

\newabbreviation{dlc}{DLC}{Double Lane Change}

\newabbreviation{ml}{ML}{Machine Learning}

\newabbreviation{cdf}{CDF}{Cumulative Distribution Function}

\newabbreviation{dare}{DARE}{Discrete Algebraic Riccati Equation}

\newabbreviation{rl}{RL}{Reinforcement Learning}

\newabbreviation{ood}{OOD}{Out of Distribution}

\newabbreviation{relu}{ReLU}{Rectified Linear Unit}

\newabbreviation{pwm}{PWM}{Pulse Width Modulation}

\newabbreviation{rnn}{RNN}{Recurrent Neural Network}

\newabbreviation{nll}{NLL}{Negative log-likelihood}

\newabbreviation{rbf}{RBF}{Radial Basis Function}

\newabbreviation{tcn}{TCN}{Temporal-convolutional}

\newabbreviation{lstm}{LSTM}{Long short-term memory}

\newabbreviation{il}{IL}{Imitation Learning}

\newabbreviation{nn}{NN}{Neural Network}

\newabbreviation{se}{SE}{Squared Exponential}

\newabbreviation{ard}{ARD}{Automatic Relevance Determination}

\newabbreviation{mle}{MLE}{Maximum Likelihood Estimation}
\newabbreviation{map}{MAP}{Maximum a posteriori}

\newabbreviation{rag}{R-Ag}{residual-agnostic}
\newabbreviation{rtg}{RTG}{Reference Trajectory Generator}

\newabbreviation{raw}{R-Aw}{residual-aware}
\newabbreviation{kde}{KDE}{Kernel Density Estimator}

\newabbreviation{cbc}{CBC}{correct-by-construction}

\newabbreviation{endo-nlp}{Endo-NLP}{NLP with endogenous residual mean constraints}

\newabbreviation{exo-nlp}{Exo-NLP}{NLP with parametrized exogenous affine residual mean}

\newglossaryentry{residual}{
    name={residual},
    description={The mismatch between next state of a discrete (or discretized) system predicted by a nominal model derived from first principles and the true measured next state (assuming no measurement error).}
}

\newglossaryentry{affine}{
    name={affine},
    description={placeholder}
}

\newglossaryentry{comb_expl}{
    name={combinatorial explosion},
    description={Refers to a situation where the number of possible combinations or outcomes of a problem grows exponentially as the problem size increases. This can quickly become overwhelming for computer algorithms, as the time and resources needed to explore every possible combination become impractical, or infeasible necessitating the use of relaxations and/or approximations to the original problem.}
}

\newglossaryentry{piecewise}{
    name={piecewise},
    description={In the context of this thesis, this term will refer to \textbf{true} underlying \glspl{residual} that vary over different regions of the state-space (or more practically, the workspace of a robotic system).}
}

\newglossaryentry{cc-mpcg}{
    name={Chance-Constrained Model Predictive Controller},
    description={
    A type of \gls{mpc} dealing with probabilistic constraints, usually used for systems involving uncertainty modelled as distributions with infinite support (\eg Gaussians).
    }
}
\newabbreviation[description={\glslink{cc-mpcg}{Chance-Constrained Model Predictive Controller}}]{cc-mpc}{CC-MPC}{Chance-Constrained Model Predictive Controller}

\newglossaryentry{hybrid}{
    name={hybrid},
    description={This shares the definition of a \gls{piecewise} system but in the context of this thesis I will use this term to indicate the learnt approximation of some underlying \gls{piecewise} \gls{residual} dynamics or to characterize an \gls{mpc} controller making use of such a model in the dynamics constraint equations.}
}

\newglossaryentry{open-loop}{
    name={open-loop},
    description={In the context of \gls{mpc}, this will be used to indicate an optimization performed at a given timestep, over a horizon into the future, of which only the first set of control inputs are applied in a receding-horizon manner.}
}

\newglossaryentry{polytope}{
    name={polytope},
    description={A n-dimensional polyhedron often used in \gls{mpc} to define constraint sets.}
}

\newglossaryentry{minlpg}{
    name={Mixed Integer Nonlinear Program},
    description={
    An extension of non-linear programming where the optimization variables can be continuous or (discrete) integer-valued. These problems can become computationally intractable to solve as the number of variables and constraints grows.
    }
}
\newabbreviation[description={\glslink{minlpg}{Mixed Integer Nonlinear Program}}]{minlp}{MINLP}{Mixed Integer Nonlinear Program}

\newglossary*{nomenclature}{Nomenclature}
\newglossaryentry{dingledorf}
{
type=nomenclature,
name=dingledorf,
description={A person of supposed average intelligence who makes incredibly brainless misjudgments}
}

\newabbreviation{aaaaz}{AAAAZ}{American Association of Amateur Astronomers and Zoologists}

\newglossary*{symbols}{List of Symbols}
\newglossaryentry{rvec}
{
name={$\mathbf{v}$},
sort={label},
type=symbols,
description={Random vector: a location in n-dimensional Cartesian space, where each dimensional component is determined by a random process}
}
\makeglossaries
\begin{document}



\title{Tractable Stochastic Hybrid Model Predictive Control using Gaussian Processes for Repetitive Tasks in Unseen Environments}

%
\author{Leroy D'Souza$^1$, Yash Vardhan Pant$^1$, Sebastian Fischmeister$^1$
\thanks{{$^1$Department of Electrical and Computer Engineering, University of Waterloo, Waterloo, Canada.}
{\tt\footnotesize \{l8dsouza, yash.pant, sebastian.fischmeister\}@uwaterloo.ca}.}%
\thanks{This work was supported in part by Magna International and the NSERC Discovery Grant.}
}

\maketitle

\begin{abstract}
Improving the predictive accuracy of a dynamics model is crucial to obtaining good control performance and safety from Model Predictive Controllers (MPC). One approach involves learning unmodelled (residual) dynamics, in addition to nominal models derived from first principles. 
Varying residual models across an environment manifest as modes of a \emph{\gls{pwr}} model that requires a) identifying how modes are distributed across the environment and b) solving a computationally intensive \gls{minlp} problem for control.
We develop an iterative mapping algorithm capable of predicting \emph{time-varying} mode distributions. 
We then develop and solve two tractable approximations of the \gls{minlp} to combine with the predictor in closed-loop to solve the overall control problem.
In simulation, we first demonstrate how the approximations improve performance by 4-18\% in comparison to the \gls{minlp} while achieving significantly lower computation times (upto 250x faster). We then demonstrate how the proposed mapping algorithm incrementally improves  controller performance (upto 3x) over multiple iterations of a trajectory tracking control task even when the mode distributions change over time. 
\end{abstract}
\section{Introduction} \label{sec:intro}
\begin{figure}[t]
\centering
\begin{tabular}{cccc}
{\includegraphics[width=0.4\textwidth, trim={0.75cm 0cm 0cm 0cm},clip]{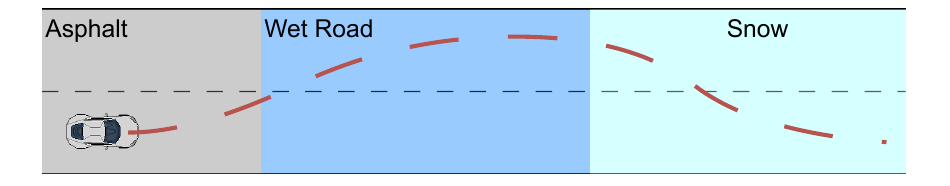}}
\end{tabular}
\caption{\small{An autonomous vehicle subject to different unmodelled (residual) dynamics depending on its position in the workspace as a result of different terrains in the environment.}}
\label{fig:dlc_motiv_exmp}
\vspace{-15pt}
\end{figure}

The accuracy of a system's dynamics model is crucial to obtaining good control performance and safety when applying \gls{mpc} to stochastic systems. To improve this accuracy, a dominant subset of data-driven control methods for systems with partially known dynamics combine a low fidelity nominal model with a learnt \emph{bayesian} model used to approximate unmodelled, or, \emph{\gls{residual}}, dynamics with uncertainty estimates~\cite{gait_gp, Hewing}. However, a single residual model can often be insufficient. 
\exmp{
Consider Fig.~\ref{fig:dlc_motiv_exmp} depicting an autonomous vehicle performing a double lane change maneuver in an environment with different terrains, viz. asphalt, a wet surface, and snow.
Assume we start off with a nominal model to predict vehicle dynamics (\eg a kinematic bicycle model). The true dynamics, however, \emph{vary} with each terrain due to different coefficients of friction. We refer to the mismatch between the nominal and true dynamics as \emph{residual} dynamics or simply, residuals. These residuals differ depending on the terrain and thus the model across this workspace is a \glsfirst{pwr} dynamics model with each individual terrain's model representing a \emph{mode} of the \gls{pwr} dynamics. Assuming that we are given residual approximations for each terrain, we still face the following challenges: \par
\noindent i) Given no information about the terrain (equivalently mode) locations (or distribution), we do not know \emph{which} residual model to use at a given point in the workspace. This prevents us from using our approximations in predicting dynamics and necessitates the use of a mapping algorithm to help identify the mode distribution from data. \par
\noindent ii) Furthermore, the mode distribution can shift over time due to robot motion or natural causes \eg additional snowfall. As a result, the previously indicated mapping algorithm must be capable of \emph{adapting} to samples generated by time-varying mode distributions in a streaming fashion.
}\label{exmp:motiv}

Recent work has demonstrated the benefits of leveraging \emph{multi-modal \mbox{(hybrid)}} learnt residual models for the purpose of control~\cite{HGPMPC_MINLP} when operating in such environments. However, they typically make the assumption that the mode distribution is \emph{known} apriori. In this paper, we consider the problem where, provided with a residual approximation to each mode of a \gls{pwr} model, the system must learn over time to operate in an environment where mode distribution is apriori unknown. We also consider that this distribution can change over time resulting in a dynamic environment (see Example~\ref{exmp:motiv}). Furthermore, since we can ``switch'' between different residual modes across the environment, this has the effect of introducing discrete variables in the predictive controller. Optimizing over such discrete variables leads to the problem of computationally \emph{intractable} \glspl{minlp} which we aim to address in this work.

\noindent\textbf{Contributions}: The contributions of this work are twofold.
    First, we propose a learning-based method to identify the distribution of \emph{modes} of residual dynamics across a potentially time-varying environment.
    Next, we propose a controller, for a system with \gls{pwr} dynamics (as described in Example~\ref{exmp:motiv}), that incorporates the mode predictions generated using the learning-based method from above. The developed controller is a computationally \emph{tractable} approximation to a baseline hybrid \gls{mpc}~\cite{HGPMPC_MINLP} (as made explicit in Section~\ref{subsec:baseline_minlp_intro}).

Through extensive simulations, we demonstrate that the proposed mapping algorithm is capable of leveraging information in bayesian residual models to build accurate maps of the \emph{time-varying} mode distribution across a previously unseen environment.
We also demonstrate that our proposed approximations still retain controller performance while significantly reducing computational time and allowing for longer \gls{mpc} horizons when compared to the baseline. We finally show that our controller achieves better control performance when compared to a baseline~\cite{multi_fric_hybrid} that does not account for mode-switching across the \gls{mpc} horizon.
The code associated with this paper can be found at {\footnotesize{\url{https://github.com/CL2-UWaterloo/AdaptiveHGPMPC_ECC25}}}.

\noindent\textbf{Related Work}: For brevity, we limit our literature review to a subset of learning-based control literature that uses bayesian models for i) hybrid model learning and  ii) improving control performance and safety of \gls{mpc} for stochastic systems. \par
\noindent\emph{Stochastic \gls{mpc}: } 
Stochastic control approaches account for distributions over disturbances, particularly ones with infinite support~\cite{SMPC_Overview}. When dealing with nonlinear functions, uncertainty propagation over a horizon can be addressed by assuming closed-form distributions (\eg Gaussians) for which tractable approximations exist~\cite{Girard}. This makes it feasible to incorporate probabilistic constraints into optimization problems~\cite{cvar_cc, Blackmore2011ChanceConstrainedOP, Hewing}. We note that addressing stochastic \gls{mpc} for hybrid systems is hard due to \gls{comb_expl} and has typically been dealt with using Markov chains in the literature~\cite{smpc_hybrid}. Our approach uses heuristics and does not provide guarantees obtained from stochastic \gls{mpc} with uni-modal models~\cite{maiworm2021online} (albeit with more restrictive assumptions).

\noindent\emph{Datasets for residual approximations: } 
Our work assumes access to datasets of residual samples to be used to train approximations to each mode's residual model. The following works do not make the above assumption.
Some work~\cite{gait_gp} uses acquisition functions for efficient exploration to learn \emph{uni-modal} bayesian models. Others provide robust guarantees when collecting new residual samples for model learning, of benefit for safety-critical systems~\cite{berkenkamp_safe_2015, koller2019learningbased}.  

\noindent\textit{Classification under concept drift:} We consider the problem of training a mode-mapping classifier over a \emph{continuous} space 
to predict the underlying \emph{time-varying} mode distribution at any point in the environment. The time-varying characteristic of the problem is referred to as concept drift in the literature~\cite{concept_drift_survey}.
This is similar to the problem of occupancy mapping, except we predict if a point is ``occupied'' by a specific mode. While this field typically deals with discretized environments, there has been recent work on extending this to continuous spaces~\cite{hilbertmapping, kernel_mapping_2020}.

Most similar to our work is~\cite{multi_fric_hybrid} which uses a history of observations to predict the current mode. This mode is assumed to be active across the entire \gls{mpc} horizon. In contrast, our mode-mapping approach is capable of accounting for switches between modes across the horizon, resulting in improved control performance as shown in Section~\ref{subsec:mode_mapping_results}.

\section{Preliminaries and Problem Setup} \label{sec:prelim}

We consider a system with state $x_k \in X \subset \mathbb{R}^n$, input $u_k \in U \subset \mathbb{R}^m$ and system dynamics of the form,
\vspace{-5pt}
\begin{equation} \label{eq:hl_sys_dyn}
x_{k+1} = f(x_k, u_k) + g(x_k, u_k) + w_k,
\end{equation}
where $f:X \times U \rightarrow X$ denotes the known \textit{nominal} model. $g: X \times U \rightarrow X$ represents the unknown \gls{pwr} dynamics with $w_k$ denoting the associated stochastic process noise term. We define $z_k := (x_k, u_k) \in Z = X \times U$. We now make the piecewise nonlinear nature of $g(x_k, u_k)$ explicit:
\noindent 
$Z$ is \emph{partitioned} by (possibly) time-varying ``regions'', $R_\text{set}(k) = \{R^r(k) \subseteq Z, \, r\in\mathbb{N}\}$ (note that $|R_\text{set}(k)|$ can change with $k$). A \emph{unique} mode of $g$ governs the dynamics in each region. 

\noindent Further, we define the following variables:
\begin{itemize}
    \item $\gset := \{\gr \mid m\in\{1,\dotsc,\numR\}\}$ to be the set of modes of $g$ (in~\eqref{eq:hl_sys_dyn}) with $\grxkuk: X \times U \rightarrow X$ where $\numR$ is the number of modes. 
    \item $R^m(k)$ as the union of all regions in $R_\text{set}(k)$ where $m$ is the active mode.
    \item $\deltartk := (1 \,\text{if}\,z_k \in R^m(k), 0\, \text{otherwise})$ as the time-varying indicator function for each mode. This function changes depending on the environment.
    \item $\wrk$ as the mode-wise process noise drawn from a Gaussian i.i.d distribution \ie $\wrk \sim \mathcal{N}\!\left(\mathbf{0},\Sigmanr\right)$. 
\end{itemize}

\noindent With this setup,~\eqref{eq:hl_sys_dyn} takes the form,
\begin{equation} \label{eq:dynamics}
x_{k+1} = f(x_k,u_k) + \sum_{m=1}^{\numR}\deltark\Big(\grxkuk + \wrk\Big).
\end{equation} 

\noindent where given that $Z$ is partitioned by regions in $R_\text{set}(k)$ each having a unique active mode, ${\sum_{m=1}^{\numR} \deltark = 1}$.

\subsection{Learning Approximations to True Dynamics}
\glsreset{gp}
Since we do not know the true dynamics~\eqref{eq:dynamics} in an unseen \emph{deployment} environment, we develop a model of the form,
\begin{subequations}
\label{eq:hl_approx_dyn}
	\begin{align}
        x_{k+1} &= f(x_k,u_k) + \hat{g}(x_k,u_k), \label{eq:hl_approx_top}\\
        \hat{g}(x_k,u_k) &= \sum_{m=1}^{\numR}\deltahatrk\Big(\ghatrxkuk\Big) \label{eq:hybrid_gp_model}
	\end{align}
\end{subequations}
where $\deltahatr(z_k), \ghatr(x_k, u_k)$ are learnt approximations to $\deltar(z_k), (\gr(x_k, u_k)+\wrk)$~\eqref{eq:dynamics} respectively. 
In our work, we assume $\gr$ is approximated by a \gls{gp} which can be seen as an extension of a multivariate Gaussian distribution to infinite sets~\cite{Girard}. Given a sample $(x_k, u_k, x_{k+1})$ collected in an environment we introduce the residual $d_k$, 
\begin{equation} \label{eq:residual}
    d_k := x_{k+1} - f(x_k, u_k).
\end{equation}
This represents the mismatch between the nominal model-predicted and true observed states (clear from~\eqref{eq:hl_approx_top}). We assume we have access to \emph{training} environment(s) where $\deltark$ is known \ie we know the mode $m$ that generated $d_k$. Other works exist that relax this assumption \cite{bemporad_parc}. 

We define $\tilde{D} := \{(x_n, u_n, d_n, m_n)\}_{n=1}^{L_{\text{train}}}$ as the dataset available to learn the \gls{gp} models. Inference with $\ghatr$ is then,
{\small
\begin{equation} \label{eq:gp_model}
    \hat{g}^m(x_k, u_k) \sim \mathcal{N}\!\left(\mu^{\hat{g}^m}(\tilde{D}^m, x_k, u_k), \Sigma^{\hat{g}^m}(\tilde{D}^m, x_k, u_k)\right)
\end{equation}
}
\noindent where {\footnotesize${\tilde{D}^m := \{(x_n, u_n, d_n) \mid (x_n, u_n, d_n, m_n) \in \tilde{D}, m_n=m\}_{n=1}^{L^m_{\text{train}}}}$} is the mode-specific dataset. Hence, given a deterministic input, inference with a \gls{gp} model yields a normal distribution whose mean and covariance is determined by the functions $\mu^{\hat{g}^m}, \Sigma^{\hat{g}^m}$ conditioned on $\tilde{D}^m$ as described further in~\cite{Girard}. We henceforth abbreviate $\mu^{\hat{g}^m}(\tilde{D}^m, x_k, u_k), \Sigma^{\hat{g}^m}(\tilde{D}^m, x_k, u_k)$ as $\mu^{\hat{g}^m}_k, \Sigma^{\hat{g}^m}_k$ respectively.

\begin{assumption} \label{as:apriori_gp}
    Bayesian \gls{gp} model approximations to each of the modes in $\gset$ are learnt from apriori provided datasets, $\{\tilde{D}^m \mid m\in\{1,\dotsc,\numR\}\}$ and denoted as $\ghatset := \{\ghatr \mid m\in\{1,\dotsc,\numR\}\}$. Any \emph{deployment} environments we consider only contain modes present in these datasets.
\end{assumption}

\noindent While $\ghatset$ can be learnt in training environments, $\deltark$ in~\eqref{eq:hl_approx_dyn} is still unknown for \emph{unseen deployment} environments (as indicated in Example~\ref{exmp:motiv}). To learn approximations $\deltahatrk$, we can collect trajectories $\{(x_k, u_k)\}_{k=1}^{L}$ in such deployment environments and use a dataset $D:=\{(x_k, u_k, d_k)\}_{k=1}^{L}$. The problem is now formally stated below.

\begin{prob}[Estimating \gls{pwr} mode distributions]
\label{prob:learn_deltahat}
Given Assumption~\ref{as:apriori_gp} holds and provided dataset $D$ generated from $\{(x_k, u_k)\}_{k=1}^{L}$ (as above), we learn mappings $\deltahat := [\deltahatone, \ldots, \deltahatmaxM]$
that takes $(x_k, u_k)$ as input and predict whether mode $m$ is active or not, \ie $\deltahatr: X \times U \rightarrow \{0, 1\}$. $\deltahat$ is then an estimator (one-hot encoding) of the active \gls{pwr} mode.
\end{prob}

\begin{rem} \label{rem:iterative_train}
Since $\delta:=[\deltaone, \ldots, \deltamaxM]$ can be time-varying, $\deltahat$ should be able to update itself when new data is available, \eg after completing a run of a tracking task (see Section~\ref{subsubsec:likelihood_prior_tradeoff}). 
Given $\ghatset$ from Assumption~\ref{as:apriori_gp} and $\deltahat$ from Problem~\ref{prob:learn_deltahat}, we now define the problem of leveraging $\hat{g}$~\eqref{eq:hl_approx_dyn} in an \gls{mpc} framework for a deployment environment.   
\end{rem}

\begin{prob}[\gls{mpc} using mode predictions]
\label{prob:cc_optimal_control}
Given nominal dynamics $f$ and a stochastic hybrid model $\hat{g}$~\eqref{eq:hl_approx_dyn}, develop a controller for the system~\eqref{eq:dynamics} that minimizes the expected cost {\small \mbox{$J=\sum_{k=1}^{N}\mathbb{E}_{w_k}(||x_k-\xref||^2_Q + ||u_k||^2_R)$}} subject to,
\begin{subequations}
\vspace{-5pt}
\label{eq:state_inp_constraints}
	\begin{align}
        \mathbb{P}(x_k \in X) \geq p_x\;\; \forall \;\; &k \geq 0 \label{eq:state_constr}\\ 
        u_k \in U\;\; \forall \;\; &k \geq 0, \label{eq:input_constr}
	\end{align}
\end{subequations}
where $N$ is the \gls{ol} lookahead horizon, $X$, $U$ are convex sets, $Q$, $R$ are positive semi-definite, positive definite matrices respectively and $p_x$ is the user-specified probability for state constraint satisfaction. The initial state at the start of each \gls{ol} optimization is known.
\end{prob}

The stochastic nature of the system over a horizon necessitates the use of probabilistic chance constraints~\eqref{eq:state_constr} to be satisfied with probability $p_x$~\cite{SMPC_Overview}. The control inputs are defined as free variables here, a choice demonstrated to work well in practice~\cite{Hewing} with alternates also being studied~\cite{Hewing2020IndirectAD}.

\subsection{Baseline \gls{minlp} controller} \label{subsec:baseline_minlp_intro}
The following \gls{hybrid} \gls{mpc} \gls{minlp} optimization was proposed in~\cite{HGPMPC_MINLP} to solve Problem~\ref{prob:cc_optimal_control} \emph{assuming apriori knowledge of $R_\text{set}(k)$} (and hence $\deltark$).
\vspace{-15pt}

{\small
\begin{subequations} \label{eq:hybrid_opt_final}
\begin{align}
\mathcal{C}_\text{MINLP}: 
\min_{u_k, \delta}& 
\; \Vert{\mu^x_N-\xrefnok_N\Vert}^2_P + \sum_{k=0}^{N-1}\Vert{\mu^x_k-\xref\Vert}^2_Q + \Vert{u_k\Vert}^2_R\label{eq:cost_minlp} \\ 
\text{s.t.} \ &\Sigma^x_{0} = \mathbf{0}_{n \times n}\;,\; \mu^x_0 = x_0, \label{eq:init_certainty} \\
& \deltarkvar \in \{0, 1\} \;\forall\; m\in\mathbb{N}_{\leq\numR},\; k\in\mathbb{N}_{\leq N} \label{eq:disc_arr}\\ 
H^m z_k \leq b^m &+ (-b^m + \vec{M})(1-\deltarkvar)\;,\; \sum_m \deltarkvar = 1 \label{eq:big_M}\\
&\mu^{\hat{g}}_{k} = \sum_{m}\deltarkvar \mughatrk \;,\; \Sigmaghatk = \sum_{m}\deltarkvar\Sigmaghatrk \label{eq:hyb_sel} \\
\mu^{x}_{k+1} &= f(\mu^x_k, u_k) + B_g \mu^{\hat{g}}_k  \label{eq:mean_dyn_minlp} \\ \label{eq:final_cov_prop_eqn}
\Sigma^x_{k+1} &= \left[\nabla \!  f(\mu_k^x,u_k) \ B_g \right] \Sigma_k {\left[\nabla \!  f(\mu_k^x,u_k)\ B_g \right]}^T\\
u_{k} \in U\;,&\;\mu^{x}_{k+1} \in \mathcal{Z}(\Sigma^x_{k+1}) \;\forall k\in\{0,\ldots,N-1\}\label{eq:shrunk_state_constraint}
\end{align}
\end{subequations} 
}

\noindent We briefly explain the key ideas behind this controller formulation, referring the reader to~\cite{HGPMPC_MINLP} for further details.~\eqref{eq:cost_minlp} is a simplification of the objective indicated in Problem~\ref{prob:cc_optimal_control}. More complex approximations often only result in marginal improvements~\cite{bradford_stochastic_2020}. \eqref{eq:init_certainty} follows from the initial state known with certainty.~\eqref{eq:disc_arr},~\eqref{eq:big_M} introduces binary variables $\deltarkvar$ and uses the big-M formulation~\cite{Morari_MPC} to embed the indicator functions $\deltark$ into the optimization.~\eqref{eq:hyb_sel} picks the mean and covariance from the active mode at timestep $k$, as determined by $z_k$. Using the output of~\eqref{eq:hyb_sel}, ~\eqref{eq:mean_dyn_minlp} yields the hybrid state mean dynamics and~\eqref{eq:final_cov_prop_eqn} uses a first order Taylor approximation for covariance dynamics across the horizon with $\Sigma_k$ denoting the joint state-input-residual covariance matrix at timestep k (as outlined in~\cite{Hewing}). $B_g$ specifies which variables in $X$ we have trained residual models for. Finally,~\eqref{eq:shrunk_state_constraint} reformulates the stochastic chance constraint~\eqref{eq:state_constr} as a deterministic constraint by shrinking as a function of the state covariance matrix $\Sigma^x_k$~\cite{Hewing}.\par

\textbf{Limitations:} In addition to requiring knowledge of the ground truth functions $\deltartk$ (\ie bypassing Problem~\ref{prob:learn_deltahat}), i) the number of $\delta^m_k$'s~\eqref{eq:disc_arr} scales with horizon length $N$ and the cardinality of $R_\text{set}(k)$ in general which significantly affects the tractability of the \gls{minlp} optimization due to combinatorial explosion, ii) the regions in $R_\text{set}(k)$ must be assumed polytopic as is for use in big-M formulation~\eqref{eq:big_M}.

\vspace{-5pt}
\section{Methodology} \label{sec:methodology}

\begin{figure*}[t]
\centering
\begin{tabular}{cccc}
{\includegraphics[width=0.8\textwidth, trim={0cm 0cm 0cm 0cm},clip]{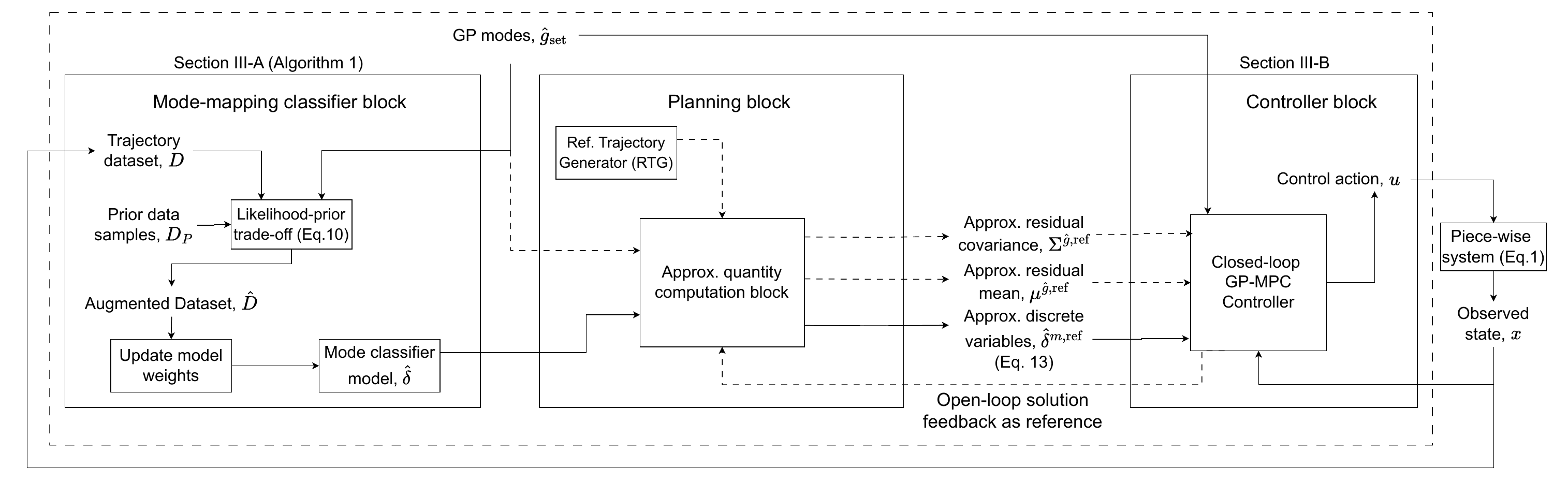}}
\end{tabular}
\caption{\small{The mode-mapping classifier block involves using a batch trajectory dataset to iteratively improve estimates of the time-varying functions $\deltartk$~\eqref{eq:dynamics} using the likelihood-prior trade-off scheme outlined in Section~\ref{subsec:mode_mapping}. The planner block uses a reference trajectory to obtain approximations to quantities which are then used to obtain \gls{nlp} approximations to  the baseline \gls{minlp} controller~\eqref{eq:hybrid_opt_final}, as further described in Section~\ref{subsec:minlp2nlp}.}}
\label{fig:blk_diag}
\vspace{-15pt}
\end{figure*}

The overall structure of the proposed control approach is shown in Fig.~\ref{fig:blk_diag}. Section~\ref{subsec:mode_mapping} outlines a solution approach to Problem~\ref{prob:learn_deltahat}. Section~\ref{subsec:minlp2nlp} presents computationally efficient approximations to the baseline \gls{minlp} controller~\eqref{eq:hybrid_opt_final} and an approach to use the solution to Problem~\ref{prob:learn_deltahat} within the proposed controllers, to solve Problem~\ref{prob:cc_optimal_control}. In doing so, we address the limitations highlighted at the end of Section~\ref{subsec:baseline_minlp_intro}.

\subsection{Learning Mode Distributions in New Environments} \label{subsec:mode_mapping}
This section elaborates on the mode-mapping classifier block seen in Fig.~\ref{fig:blk_diag}. We introduce $\ydk, \ygk$ such that $\deltahatrk \equiv \deltahatrydk$, $\ghatrxkuk \equiv \ghatrygk$. 
Thus, from any trajectory $\{(z_k)\}_{k=1}^{L}$, we define the function $\hybgpinpgen$ as, $\hybgpinpgen(\{z_k\}_{k=1}^{L}) := \{(\ygk, \ydk, \dk)\}_{k=1}^{L-1}$, with $\dk$ as in~\eqref{eq:residual}. We aim to \emph{iteratively} (as indicated by Remark~\ref{rem:iterative_train}) train a mode-mapping classifier, $\hat{\delta}$, that outputs a probability distribution over modes corresponding to $\ydk$, \ie $\hat{\delta}(\ydk) := [\deltahatnumfn{1}{\ydk},\ldots,\deltahatnumfn{\numR}{\ydk}]^T$ and approximates $\delta(\ydk)$ in~\eqref{eq:dynamics}. 

The proposed solution is summarized in Algorithm~\ref{alg:modemapping}. Line~\ref{algline:likelihood} uses the models $\ghatset$ and inputs $\ygk$ to compute the likelihood that a particular mode generated a measured residual sample $d_k$ (Section~\ref{subsubsec:likelihood_comp}). Line~\ref{algline:prior} uses the current classifier $\hat{\delta}$ to predict prior probabilities over the modes. Line~\ref{algline:tradeoff} uses confidence estimates in the prior probabilities to trade-off likelihood and prior (Section~\ref{subsubsec:likelihood_prior_tradeoff}) with Line~\ref{algline:retrain} fine-tuning $\hat{\delta}$ on the new batch of training labels. We use the SIREN network architecture~\cite{siren} for our mapping predictor.

\begin{algorithm}
\linespread{1.1}\selectfont
\caption{Iterative mode mapping}
\label{alg:modemapping}
\hspace*{\algorithmicindent} \textbf{Input: } \gls{gp} modes $\ghatset$, trajectory data $D:=\{z_k\}_{k=1}^{L}$, current classifier $\hat{\delta}$, Prior dataset $\prevDS$\\
\begin{algorithmic}[1]
    \State $\{(\ygk, \ydk, \dk)\}_{k=1}^{L} \gets \hybgpinpgen(\{z_k\}_{k=1}^{L})$
    \State $\lmk \gets \texttt{ComputeLikelihoods}(\ghatset, \ygk)$ \Comment{As in~\eqref{eq:sample_prob_hyb}} \label{algline:likelihood}
    \State $\priormk \gets \texttt{ComputePriors}(\mapapprox, \ydk)$ \label{algline:prior}
    \LineComment{Likelihood-prior trade-off as in~\eqref{eq:tradeoff}}
    \State $\yk \gets \texttt{ComputePosteriors}(\lmk, \priormk, \prevDS)$ ~\label{algline:tradeoff}
    \State $\hat{D} \gets \{{\ydk, \yk}\}_{k=1}^{L}$ \Comment{Augmented Dataset}
    \State $\mapapprox \gets \texttt{Retrain}(\hat{D})$ \label{algline:retrain}  
    \State $\prevDS \gets \prevDS \cup \hat{D}$ \label{algline:updatepriorDS}
\end{algorithmic}
\end{algorithm}

\subsubsection{Likelihood computation using $\ghatset$ and $\dk$} \label{subsubsec:likelihood_comp}
\begin{figure}[h]
\centering
\includegraphics[width=0.5\textwidth, trim={0cm 0.5cm 0cm 0.7cm},clip]{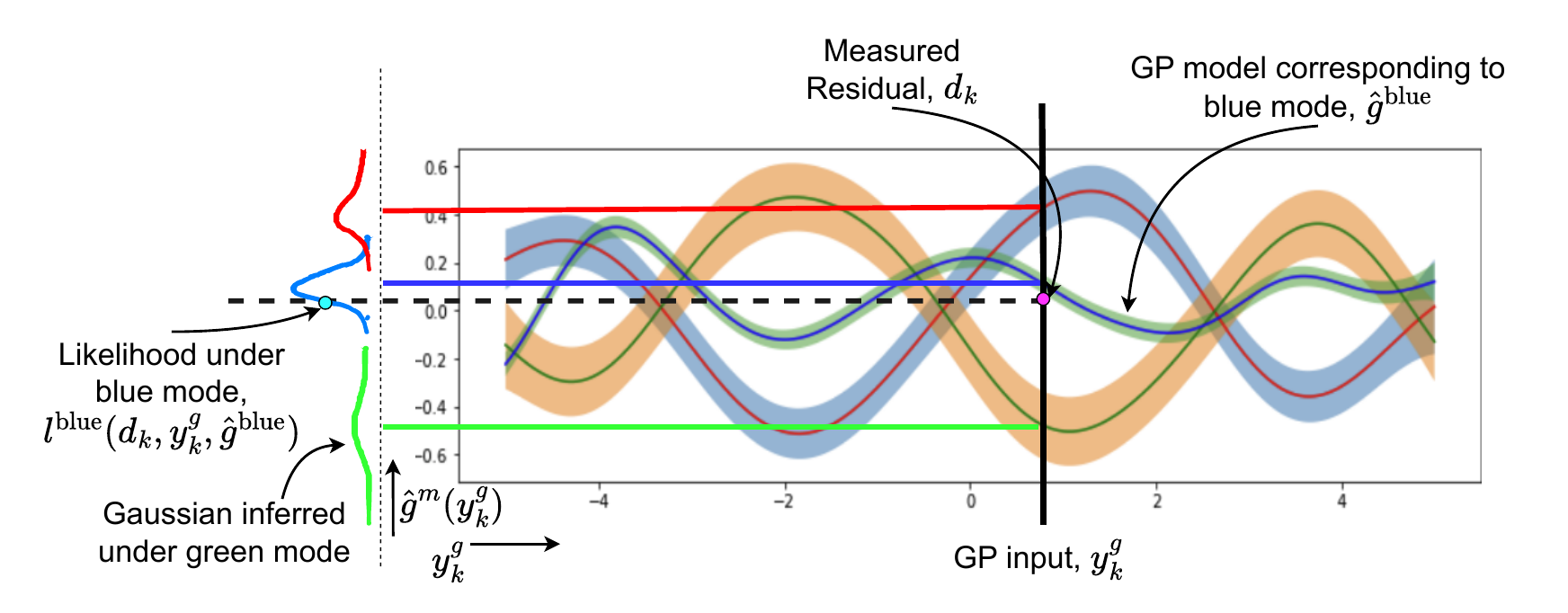}
\caption{\small{A visualization of how each \gls{gp} model in $\ghatset$ infers a Gaussian given a deterministic input. The likelihood of the true measured residual under each mode's Gaussian can be computed.}}
\label{fig:likelihood_comp_viz}
\vspace{-5pt}
\end{figure}

Since we know that the \gls{gp} output for a deterministic input is a Gaussian (as further visualized in Figure~\ref{fig:likelihood_comp_viz}), the likelihood that a residual sample, $\dk$, was generated from a given mode, $\ghatr$ can be computed as,

\begin{subequations} \label{eq:sample_prob_hyb} 
\footnotesize
\begin{align}
\lrdghat &:= \mathbb{P}(\dk \mid \ygk, \Rzkr)\\
= \frac{1}{(2\pi)^{n_d/2}\det(\Sigmaghatrk)^{1/2}} \text{exp} &\left(-\frac{1}{2}(d_k - \mughatrk)^T {\Sigmaghatrk}^{-1} (d_k - \mughatrk) \right) \nonumber
\end{align}
\end{subequations}
\normalsize
where $n_d$ is the size of the residual vector, $\Rzkr \iff \deltartk=1$. As we do not know the functions $\deltartk$, we simply iterate over all possible modes to compute the likelihood of $d_k$ under each mode.

\subsubsection{Posterior probability computation} 
Since we design an iterative mapping algorithm, we would like to be able to use the predictions from $\hat{\delta}$ as prior probabilities, in addition to the computed likelihood~\eqref{eq:sample_prob_hyb}.
We recall the following basic rules from probability theory, (i) conditional probability, $\mathbb{P}(X\mid Y) \mathbb{P}(Y) = \mathbb{P}(X, Y)$, (ii) chain rule, $\mathbb{P}(X, Y, Z) = \mathbb{P}(X\mid Y, Z)\mathbb{P}(Y\mid Z)$.
Since we desire a distribution over modes, any terms independent of $m$ are proportionality constants.

\begin{figure}[h]
\centering
\includegraphics[width=0.3\textwidth, trim={0cm 0cm 0cm 0cm},clip]{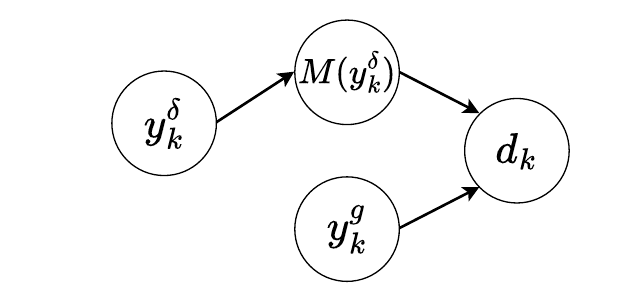}
\caption{\small{A directed graph showing dependence relations between the variables affecting the residual magnitude. Note $M(y^\delta_k) \equiv M(z_k)$.}}
\label{fig:dependence_diagram}
\vspace{-10pt}
\end{figure}

\footnotesize
\begin{subequations} 
\begin{align}
    \mathbb{P} (\dk \mid \ygk, \Rzkr) &=\frac{\mathbb{P} (\Rzkr \mid \ygk, \dk) \mathbb{P}(\dk \mid \ygk)}{\mathbb{P} (\Rzkr, \ygk)} \\
    &\propto \frac{\mathbb{P} (\Rzkr \mid \ygk, \dk)}{\mathbb{P} (\Rzkr) \cdot \mathbb{P}(\ygk)}  \label{eq:Rzk_cond_indep} \\
    \mathbb{P} (\Rzkr \mid \ygk, \dk) &\propto \mathbb{P} (\dk \mid \ygk, \Rzkr)\cdot \mathbb{P} (\Rzkr) \nonumber\\
    &= \lrdghat \cdot \deltahatrydk \label{eq:posterior_rule}
\end{align}
\end{subequations}
\normalsize

\noindent\eqref{eq:Rzk_cond_indep} follows since $\Rzkr$ is conditionally independent of $\ygk$ if $\dk$ is not observed (Fig.~\ref{fig:dependence_diagram}),~\eqref{eq:posterior_rule} follows from~\eqref{eq:sample_prob_hyb}.

\subsubsection{Likelihood-prior trade-off for adaptation} \label{subsubsec:likelihood_prior_tradeoff}
\begin{figure}[h]
\centering
\includegraphics[width=0.45\textwidth, trim={0cm 0cm 0cm 0cm},clip]{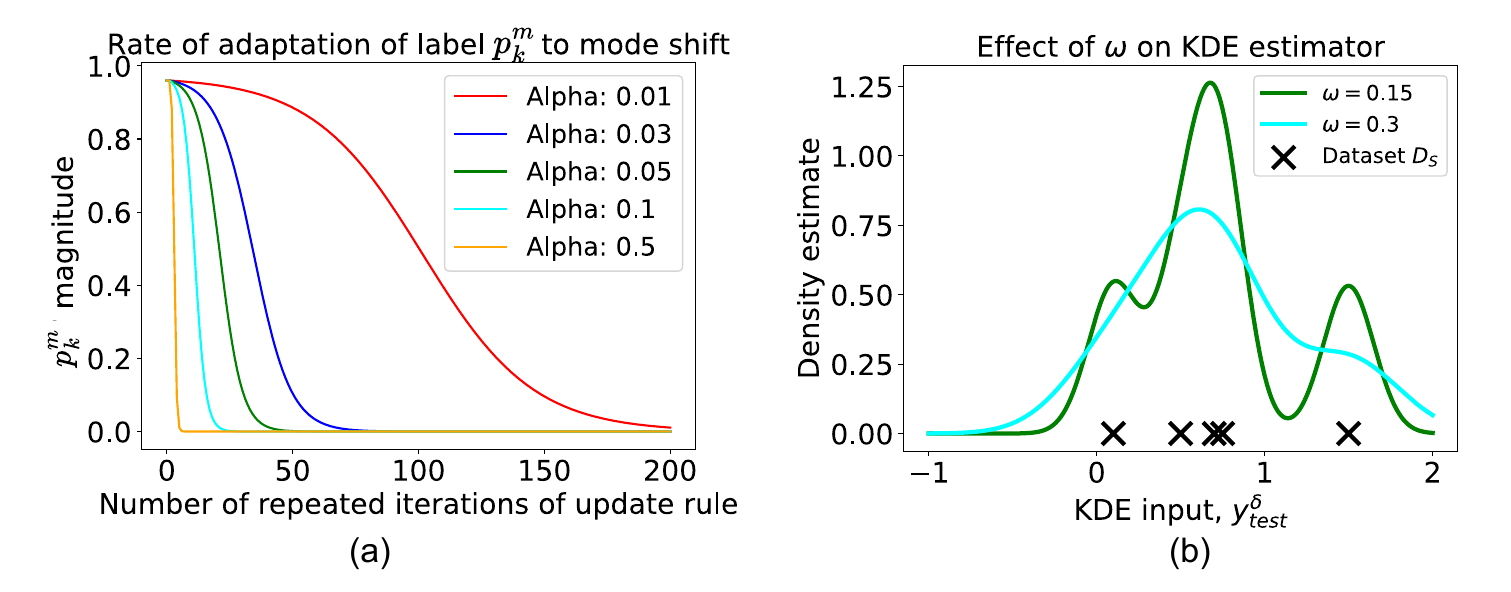}
\caption{\small{(a) For a prior of 0.95 and likelihood of 0.05, this graph shows that smaller $\alpha_k(\ydk)$ results in slower adaptation of the posterior to the change in the likelihood (\ie mode distribution). (b) A demonstration of how larger $\kdebw$ yields a smoother $\kdeest(\cdot)$ function.}}
\label{fig:tradeoff_basics}
\vspace{-10pt}
\end{figure}

The mode-mapping predictor, $\hat{\delta}$, ideally adapts quickly to information in new samples in parts of the environment that have not been seen before (low prior data density). It must also adapt to changes in the underlying mode distribution over time. 
With this in mind, we modify the update rule~\eqref{eq:posterior_rule} to trade off the prior and likelihood terms as follows,
\begin{equation} \label{eq:tradeoff}
    \ymk := (\lrdghat)^{\alpha(\ydk)}) \cdot \deltahatrydk / c
\end{equation}
where $\alpha(\cdot)$ is a trade-off function, $c$ is a normalization constant so that $\yk = \{\yonek,\ldots,\yMk\}$ sums to 1.
Due to normalization in~\eqref{eq:tradeoff}, a higher $\alpha(\ydk)$ has the effect of boosting trust in the likelihood predictions which helps deal with inaccurate prior predictions. A lower $\alpha(\ydk)$ trusts the prior more and can be used to specify the rate of adaptation to changes in the mode distribution as seen in Fig.~\ref{fig:tradeoff_basics}(a).
This approach can be viewed as similar to others in the literature that use a \gls{map} approach to address overconfidence in \glspl{nn}~\cite{overconfidence}. 

To obtain $\alpha(\ydk)$ we first generate a confidence score in the $\hat{\delta}$ prediction using a \gls{kde}, similar to distance-based confidence prediction~\cite{conf_pred}. The \gls{kde} parameter, $\kdebw$, affects the shape of the estimate as shown in Fig.~\ref{fig:tradeoff_basics}(b). Given a prior dataset $D_P = \{\ydkprior\}_{k_P=0}^{N_P}$ of size $N_P$, the \gls{kde} estimate at a new point, $\ydk$, is computed as, 

{\footnotesize
\begin{equation} \label{eq:kde_est}
    \kdeest(\ydk) = \frac{1}{\kdebw N_P}  \sum_{k_P=0}^{N_P}\left(\frac{1}{\sqrt{2\pi}}\text{exp}(\ydkprior - \ydk)^2 / (2\kdebw^2)\right)
\end{equation}}

\begin{figure}[h]
\centering
\includegraphics[width=0.45\textwidth, trim={0cm 0cm 0cm 0cm},clip]{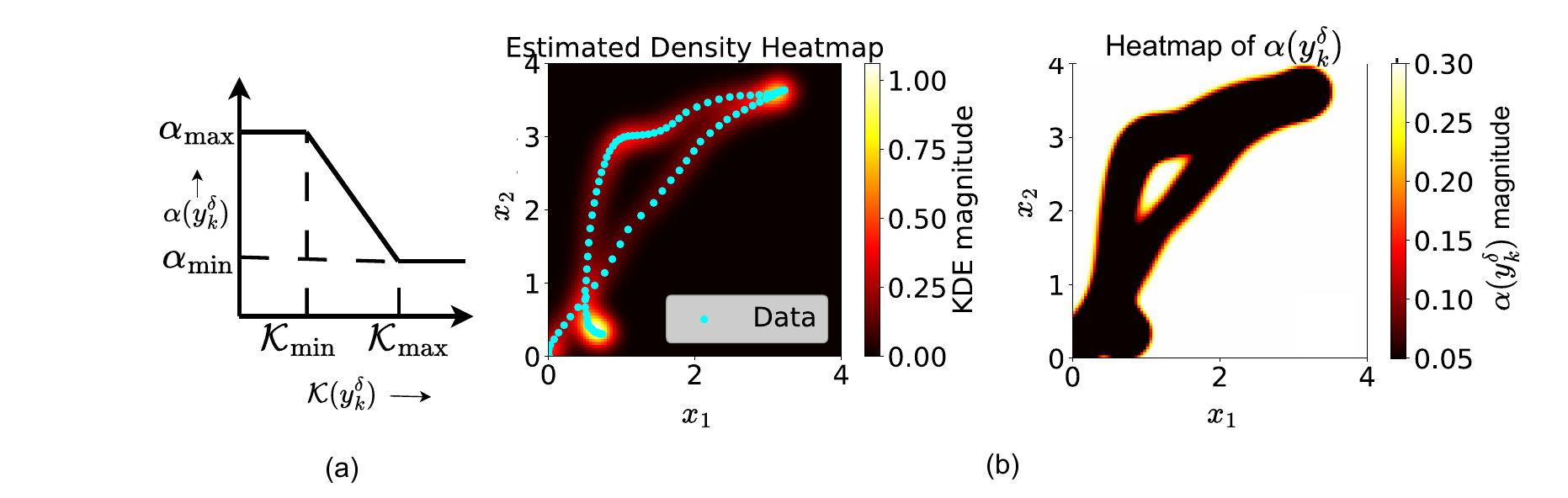}
\caption{\small{(a) Clamped linear function converting $\kdeest(\ydk)$ to $\alpha(\ydk)$, (b) Visualization of KDE and $\alpha(\cdot)$ heatmaps on an example dataset.}}
\label{fig:tradeoff_viz}
\vspace{-10pt}
\end{figure}

$\kdeest(\ydk)$ is now converted to $\alpha(\ydk)$ by a linear transformation parametrized by hyperparameters $\kdemax, \kdemin, \alphamax, \alphamin$ as depicted in Fig.~\ref{fig:tradeoff_viz}(a). A higher value for $\kdeest(\ydk)$ yields a comparatively lower $\alpha(\ydk)$, placing more trust in the prior since data from previously collected trajectories has passed close to $\ydk$. The effect of $\alphamax, \alphamin$ are motivated by prior discussion on the meaning of $\alpha(\ydk)$ and further exemplified in Section~\ref{subsec:mode_mapping_results}. An example of this procedure is seen in Fig.~\ref{fig:tradeoff_viz}(b). The classifier $\hat{\delta}$ can now be fine-tuned on these labels, $\yk$. It is of note that $D_P$ changes at each iteration of Algorithm~\ref{alg:modemapping} (Line~\ref{algline:updatepriorDS}), hence changing $\alpha(\cdot)$ itself. 

\subsection{Controller design} \label{subsec:minlp2nlp}
\noindent In this section, we address the two limitations of the baseline outlined in Section~\ref{subsec:baseline_minlp_intro}. 
We initially assume the ground truth functions $\deltarset$ are known. Given the \gls{minlp} controller~\eqref{eq:hybrid_opt_final}, we derive \gls{nlp} approximations, improving computational tractability while retaining control performance.
Next, we incorporate the hybrid \gls{gp} model~\eqref{eq:hybrid_gp_model} obtained using the classifier $\hat{\delta}$ trained as in Section~\ref{subsec:mode_mapping}. This relaxes the controller's assumption that $\deltarset$ is known.

To obtain a tractable \gls{nlp} solution from an intractable \gls{minlp}, we convert several variables in~\eqref{eq:hybrid_opt_final} to parametric quantities that are generated from a reference trajectory, $\zrefkoverN$ (where $N$ is the \gls{ol} horizon), prior to the optimization. For the rest of this paper, we consider the $\zrefkoverN$  to be the \gls{ol} solution generated offline by an optimization that uses the nominal model, $f$~\eqref{eq:hl_sys_dyn}. This is the output from the \gls{rtg} block in Fig.~\ref{fig:blk_diag}.
Alternatively, the online \gls{ol} solution generated by the last run of the optimization could also be used in closed-loop (realized via a feedback loop from the \gls{mpc} controller block to the approximate quantity computation block in Fig.~\ref{fig:blk_diag}).

The intractability of the \gls{minlp}~\eqref{eq:hybrid_opt_final} mainly stems from the inclusion of binary optimization variables~\eqref{eq:disc_arr}. Instead, \emph{parametrizing} these variables as $\deltarrefk = 1 \iff z^{\text{ref}}_k \in R^m$ results in a more tractable \gls{nlp} optimization since $z^{\text{ref}}_k$ is known prior to running the optimization and hence so is $\deltarrefk$. Since~\eqref{eq:cost_minlp} no longer optimizes over binary variables, $\deltarkvar$,~\eqref{eq:disc_arr} and~\eqref{eq:big_M} can be dropped from the formulation. $\deltarrefk$ replaces $\delta^m_k$ in~\eqref{eq:hyb_sel}. Similar to Section~\ref{subsec:mode_mapping}, we define $\ygrefk$, obtained from $\zrefk$ which can be an input to $\ghatr$. With this in place, we now describe the two incremental approximations to~\eqref{eq:hybrid_opt_final} for which we demonstrate results in Section~\ref{subsec:lti_results_comparison}.

$\cnlpendo$: This controller computes the shrunk constraint sets~\eqref{eq:shrunk_state_constraint} offline. Approximate residual covariances are computed as, $\Sigmaghatrefk = \sum_{m}\deltarrefk\Sigmaghatrefk$ with $\Sigmaghatrefk = \Sigma^{\hat{g}_m}(\tilde{D}, \xref, \uref)$ using~\eqref{eq:gp_model}. Thus,~\eqref{eq:final_cov_prop_eqn} can be dropped from the optimization, computing approximate state covariance matrices, $\Sigmaxrefk$, using the procedure outlined in~\cite{HGPMPC_MINLP} as a function of $\Sigmaghatrefk$. The approximate shrunk constraint sets, $\mathcal{Z}^{\text{ref}}(\Sigmaxrefkplus)$ replace $\mathcal{Z}(\Sigma^x_{k+1})$ in~\eqref{eq:shrunk_state_constraint}.\par
$\cnlpexo$: Finally, we can treat the \gls{gp} as completely external to the optimization by computing a \emph{parametrized} hybrid residual mean $\mughatrefk$ from $\ygrefk$ to replace the optimization variable $\mu^{\hat{g}}_{k}$. Hence,~\eqref{eq:mean_dyn_minlp} reduces to a nominal dynamics model with a parametric affine term. For clarity, this optimization is summarized below,

{\small
\begin{subequations} \vspace{-8pt}\label{eq:hybrid_opt_exonlp_final}
\begin{align*}
\cnlpexo:  \min_{u_k}& 
\; \Vert{x_N-\xrefnok_N\Vert}^2_P + \sum_{k=0}^{N-1}\Vert{\mu^x_k-\xref\Vert}^2_Q + \Vert{u_k\Vert}^2_R\\
\text{s.t.} \ &\Sigma^x_{0} = \mathbf{0}_{n \times n}\;,\; \mu^x_0 = x_0, \nonumber \\ 
&\mu^{x}_{k+1} = f(\mu^x_k, u_k) + B_g \mughatrefk, \\
u_{k} \in U\;,&\;\mu^{x}_{k+1} \in \mathcal{Z}(\Sigmaxrefkplus) \;\forall k\in\{0,\ldots,N-1\}\nonumber
\end{align*}
\end{subequations}} 

To relax the assumption that $\deltarset$ are known, we first recall that ${\sum_{m=1}^{\numR} \deltarrefk = 1}$ (for reasons outlined in Section~\ref{sec:prelim}) and hence 
$\delta^{\text{ref}}_k = [\deltaonerefk,\ldots,\deltaMrefk]$ is a one-hot vector. In contrast, $\hat{\delta}$ (as in Algorithm~\ref{alg:modemapping}), outputs a \emph{soft-label} vector, $\hat{\delta}(\ydrefk)$, for a given input $\ydrefk$. Hence we define,
{\footnotesize
\begin{equation} \label{eq:deltahat_ref_gen}
\deltahatrrefk =
    \begin{cases}
      1, & \text{if} \;m = \text{argmax}(\hat{\delta}(\ydrefk)) \\
      0, & \text{otherwise}
    \end{cases}
    ,\;\forall k \in \{0, \ldots, N\}
\end{equation}}
which converts soft labels to one-hot vectors. Hence $\hat{\delta}^{\text{ref}}_k = [\deltahatonerefk,\ldots,\deltahatMrefk]$ satisfies the requirement outlined in Problem~\ref{prob:learn_deltahat}. $\deltahatrrefk$ replaces $\deltarrefk$ in $\cnlpendo, \cnlpexo$ with the rest of the construction unchanged. These controllers can now take the place of the controller block in Fig.~\ref{fig:blk_diag}.

\section{Simulation studies}\label{sec:results}

\vspace{-4pt}
The proposed controllers are implemented in Python using Casadi~\cite{Casadi}. The BONMIN~\cite{BONMIN} solver is used for the \gls{minlp} controller experiments with IPOPT~\cite{IPOPT} being used for the \gls{nlp} controller experiments. 
\Gls{cl} cost is used as a metric to compare the controllers and is defined as,

{\small
\begin{equation*} \label{eq:cl_cost}
\text{C.L. cost} = ||x_T-\xrefT||^2_Q + \sum_{k=0}^{T-1} (||x_k-\xrefk||^2_Q + ||u_k||^2_R)
\end{equation*}} 
\noindent where $T$ is the number of simulation steps, $\{\xrefk\}_{k=0}^{T}$ is the state trajectory to track, $Q$, $R$ are cost matrices defined in~\eqref{eq:hybrid_opt_final}. For the stochastic system we define $\clcostmean, \clcoststddev$ as the empirical mean and standard deviation of the cost over multiple runs.

Section~\ref{subsec:lti_results_comparison} compares the \gls{nlp} controllers $\cnlpendo, \cnlpexo$ to the \gls{minlp} controller baseline~\eqref{eq:hybrid_opt_final} when the mode distribution is \emph{known} (for a fair comparison against the \gls{minlp}). Section~\ref{subsec:mode_mapping_results} demonstrates the working of the $\cnlpexo$ controller, when combined with the mode mapping predictor $\hat{\delta}$, under unknown mode distributions (as described in Section~\ref{subsec:minlp2nlp}). This is compared against a baseline similar to the approach outlined in~\cite{multi_fric_hybrid}. 

\subsection{Planar LTI System Experiments} 
\label{subsec:lti_results_comparison}

\begin{figure}[t]
\centering
\begin{tabular}{cccc}
{\includegraphics[width=0.45\textwidth, trim={0cm 0cm 0cm 0cm},clip]{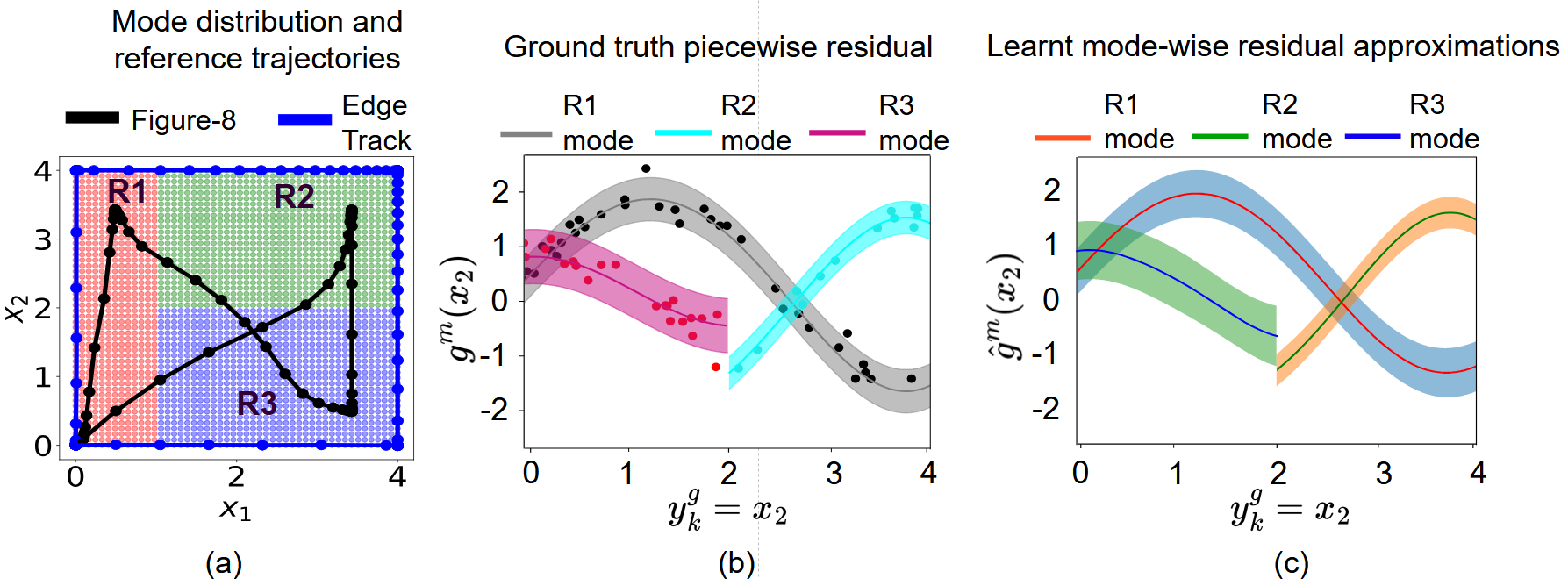}}
\end{tabular}
\vspace{-8pt}
\caption{\small{(a) Regions $R_1, R_2, R_3$ across $x_1-x_2$ space along with reference trajectories for experiments. (b) Ground truth \gls{pwr} dynamics models and residual data samples. (c) Learnt mode-wise residual approximations. $R_2$ and $R_3$ modes are truncated based on the $x_2$ bounds that define them based on the choice of $\ygk, \ydk$.}}
\label{fig:true_func_lti}
\vspace{-10pt}
\end{figure}

We consider a planar \gls{lti} system with nominal dynamics of the form $\dot{x} = Ax + Bu$ where,
\begin{subequations}
    \begin{gather}
    A = \begin{bmatrix}
    1 & 0\\
    0 & -1
    \end{bmatrix}, \quad
    B = \begin{bmatrix}
    1 & 0\\
    0 & 1
    \end{bmatrix}.
    \end{gather}
\end{subequations}
The state constraint set $X$ has $0 \leq x_1 \leq 4$, $-0.05 \leq x_2 \leq 4$ with $\ydk=(x_{1, k}, x_{2, k})$, partitioned as shown in Figure~\ref{fig:true_func_lti}(a). We choose $\gin=x_2$ with the nonlinear residual dynamics affecting the $x_1$ dynamics (\ie ${B_g = [1\,,\, 0]^T}$ in~\eqref{eq:mean_dyn_minlp}) as shown in Figure~\ref{fig:true_func_lti}(b) with $\Sigma^1=0.2, \Sigma^2=0.15, \Sigma^3=0.25$ (for $\Sigma^m$ defining the process noise distribution in~\eqref{eq:dynamics}). We let $U = \{ u \mid -5 \leq u_{i} \leq 5 \;\forall\; i \in \{1, 2\}\}$. $R=0.01I_{2 \times 2}$, $Q=50I_{2 \times 2}$ are used for the online controllers~\eqref{eq:cost_minlp}. 

We demonstrate our methodology on two tasks, a Figure-8 trajectory and a trajectory tracking the edge of the workspace as seen in Figure~\ref{fig:true_func_lti}(a). 

\par
\subsubsection{Figure-8 Tracking} \label{subsubsec:figure8lti}
\begin{figure}[t]
\centering
\begin{tabular}{cccc}
{\includegraphics[width=0.4\textwidth, trim={0cm 0.25cm 0cm 0cm},clip]{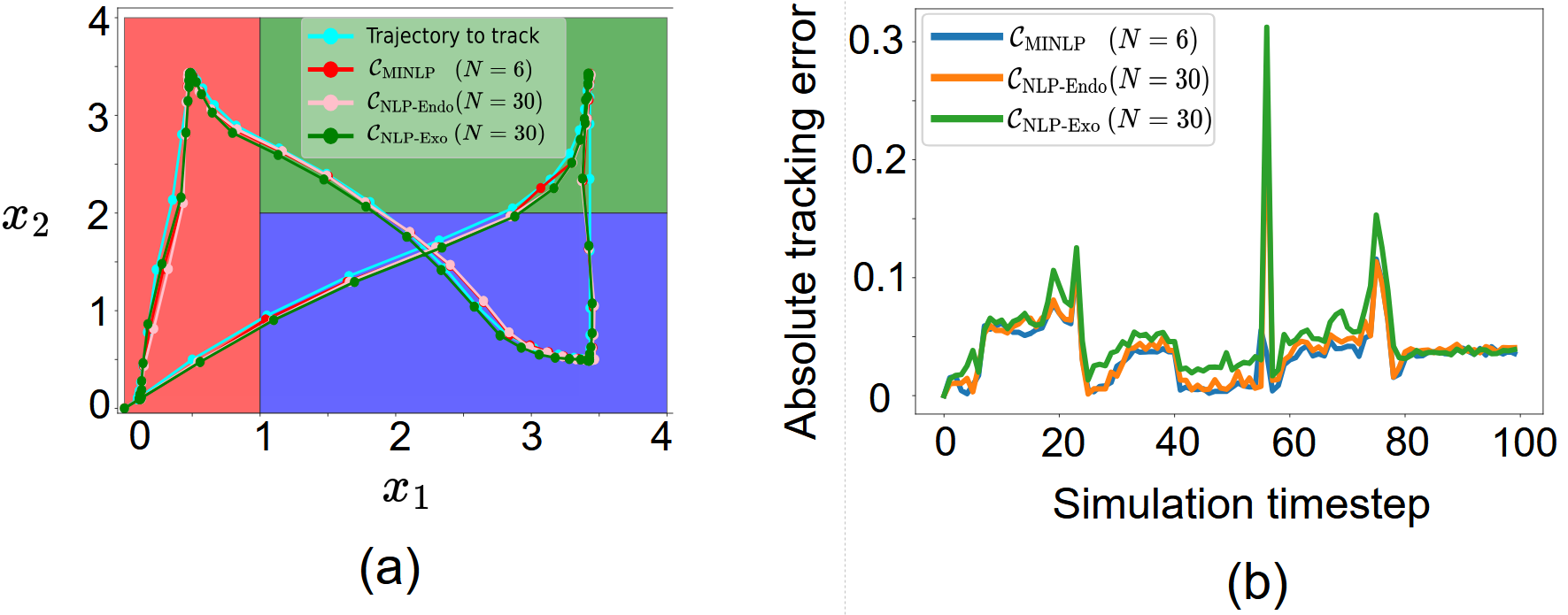}}
\end{tabular}
\caption{\small{Both plots demonstrate comparisons against 3 controllers viz., $\cminlp$ with $N=6$, $\cnlpendo$ and $\cnlpexo$ with $N=30$ for the setup as described in Fig~\ref{fig:true_func_lti}. (a) \gls{cl} trajectory comparison for Figure-8 tracking task. (b) Plot of absolute error as a function of the simulation timestep during the boundary tracking task.}}
\label{fig:cl_lti_qualit}
\end{figure}

The trajectory to track for this test is well within the state constraint boundaries as shown in Figure~\ref{fig:true_func_lti}(a). The \gls{cl} cost statistics over 50 simulation runs on this task are summarized in Table~\ref{table:int_tracking_lti} with an example visualization in Fig.~\ref{fig:cl_lti_qualit}(a). \par
\noindent\textbf{Discussion: } 
Set shrinking \emph{within} the $\cminlp$ controller optimization caused the average \gls{ol} computation time to increase significantly to $\sim$22.74s (compared to $\sim$2s for $N$=6 in Table~\ref{table:int_tracking_lti}) while obtaining similar average cost. As a result, the $\cminlp$ results use the same parametrized set shrinking approach as in the $\cnlpendo$ and $\cnlpexo$ controllers even though the $\deltark$'s are optimization variables~\eqref{eq:disc_arr}.

Going from a horizon of $N=6$ to $N=10$ for $\cminlp$ more than triples the average computation time per \gls{ol} optimization solve due to the larger number of discrete variables. The $\cnlpendo$ and $\cnlpexo$ controllers deal with significantly longer \gls{ol} horizons while demonstrating much lower computation times per \gls{ol} solve. The general trend in Table~\ref{table:int_tracking_lti} shows that computation time from $\cminlp \rightarrow \cnlpendo \rightarrow \cnlpexo$ decreases at a significantly faster rate than the corresponding increase in cost that results from the incremental approximations outlined in Section~\ref{subsec:minlp2nlp}. 

\begin{table}[htb!] \vspace{-5pt}
\centering
\begin{tabular}{rrlrl} 
 \toprule
 Controller & \multicolumn{2}{c}{$\clcostmean$} & \multicolumn{2}{c}{Avg. \gls{ol} comp. time (ms) }\\
 \cmidrule(lr){1-1} \cmidrule(lr){2-3} \cmidrule(lr){4-5}
$\cminlp$~\eqref{eq:hybrid_opt_final} (N=10) & 26.61 & (x1.01) & 6430 & (x254.7)\\ 
$\cminlp$ (N=6) & \textbf{26.39} & & 2002 & (x79.3)\\ 
$\cnlpendo$ (N=30) & 27.40 & (x1.04) & 84 & (x3.3)\\ 
$\cnlpexo$ (N=30) & 30.04 & (x1.14) & \textbf{25} & \\ 
\bottomrule
\end{tabular}
\caption{\small{\gls{cl} cost and average \gls{ol} optimization computation time trends, over 50 \gls{cl} simulation runs, for $\cminlp$ controllers with $N=\{6, 10\}$ and $\cnlpendo, \cnlpexo$ controllers with $N=30$ tracking a Figure-8 trajectory. $\clcoststddev$ was orders of magnitude lower than $\clcostmean$ across all controllers and hence not reported.}}
\label{table:int_tracking_lti}
\vspace{-10pt}
\end{table}

\subsubsection{Boundary tracking task} \label{subsubsec:boundarylti}
For this experiment, the trajectory to track operates at the edges of the state space as visualized in Figure~\ref{fig:true_func_lti}(a). With this test, we aim to test whether the controller meets a user-specified constraints satisfaction threshold, $p_x$ as defined in Problem~\ref{prob:cc_optimal_control}. Our tests use $p_x = \{0.99, 0.9\}$ with ideal average violations of 1, 10 respectively for a simulation length of 100 timesteps. We define the absolute tracking error as ${\sum_j |x_{j, k} - x^{\text{ref}}_{j, k}|}$ with $x_{j, k}$ being the $j^{th}$ element of the state, $x$, at timestep $k$. \par

\noindent\textbf{Discussion: } 
Once again, Fig.~\ref{fig:cl_lti_qualit}(b) shows that $\cnlpexo$ consistently has a higher absolute tracking error across the trajectory while the plot for $\cnlpendo$ remains comparatively close to that for $\cminlp$. Table~\ref{table:safety_viol_lti} validates that $p_x=0.9$ results in a less conservative controller than $p_x=0.99$ yielding a lower cost, as is intuitive for this specific task. While the behaviour of the controllers is generally conservative (expected due to conservative set shrinking approach~\cite{Hewing}), neglecting to include shrinking can lead to high violations. This is seen in Table~\ref{table:safety_viol_lti} where running $\cnlpendo$ without set shrinking results in significantly lower constraint satisfaction than desired with $p_x=\{0.99, 0.9\}$. Table~\ref{table:safety_viol_lti} displays the same cost trends across the controllers as seen in Table~\ref{table:int_tracking_lti}.

\begin{table}[htb!] 
\centering
\begin{tabular}{crlr} 
 \toprule
 Controller & \multicolumn{2}{c}{$\clcostmean$} & Avg. violations per run\\
 \cmidrule(lr){1-1} \cmidrule(lr){2-3} \cmidrule(lr){4-4}
$\cnlpendo$ w/o shrinking & \textcolor{red}{\bf 32.68} & & \textcolor{red}{\bf 32.16} \\ \cmidrule(lr){1-1} \cmidrule(lr){2-3} \cmidrule(lr){4-4}
$\cminlp$ ($p_x=0.99$) & \bf 43.42 & & 0 \\ 
$\cnlpendo$ ($p_x=0.99$) & 46.87 & (x1.08) & 0\\ 
$\cnlpexo$ ($p_x=0.99$) & 50.23 & (x1.16) & 0\\ \cmidrule(lr){1-1} \cmidrule(lr){2-3} \cmidrule(lr){4-4}
$\cminlp$ ($p_x=0.9$) & \bf 36.35 & & 0.05 \\ 
$\cnlpendo$ ($p_x=0.9$) & 39.36 & (x1.08) & 1.40\\ 
$\cnlpexo$ ($p_x=0.9$) & 42.49  & (x1.17) & 0.10\\ 
\bottomrule
\end{tabular}
\caption{\small{\gls{cl} cost for $\cminlp$ controller with $N=6$ and $\cnlpendo, \cnlpexo$ controllers with $N=30$ tracking a trajectory near the state constraint boundary. The $\cnlpendo$ controller without shrinking is highlighted in red due as it violates $p_x \in \{0.99, 0.9\}$. Results were computed over 30 simulation runs for each controller.}}
\label{table:safety_viol_lti}
\end{table}

\subsection{Adaptive Mode-mapping on a 2D quadrotor} \label{subsec:mode_mapping_results}
Having established the benefits of using the $\cnlpexo$ controller~\eqref{eq:hybrid_opt_exonlp_final} in Section~\ref{subsec:lti_results_comparison}, we will demonstrate the results of our mode-mapping algorithm~\ref{alg:modemapping} for this controller using the approximations given by~\eqref{eq:deltahat_ref_gen}.
The baseline controller in this section will \emph{fix} the value of the $\deltahatrrefk$ based on $\ydrefzero$ \emph{alone} for each \gls{ol} optimization (similar to the approach used in~\cite{multi_fric_hybrid}). Hence, for the baseline,~\eqref{eq:deltahat_ref_gen} now changes to,

{\footnotesize
\begin{equation} \label{eq:fixed_delta_param}
\deltahatrrefk =
    \begin{cases}
      1, & \text{if} \;m = \text{argmax}(\hat{\delta}(\ydrefzero)) \\
      0, & \text{otherwise}
    \end{cases}
    ,\;\forall k \in \{0, \ldots, N\}
\end{equation}}

\begin{figure}[t]
\centering
\begin{tabular}{cccc}
{\includegraphics[width=0.45\textwidth, trim={0cm 0cm 0cm 0cm},clip]{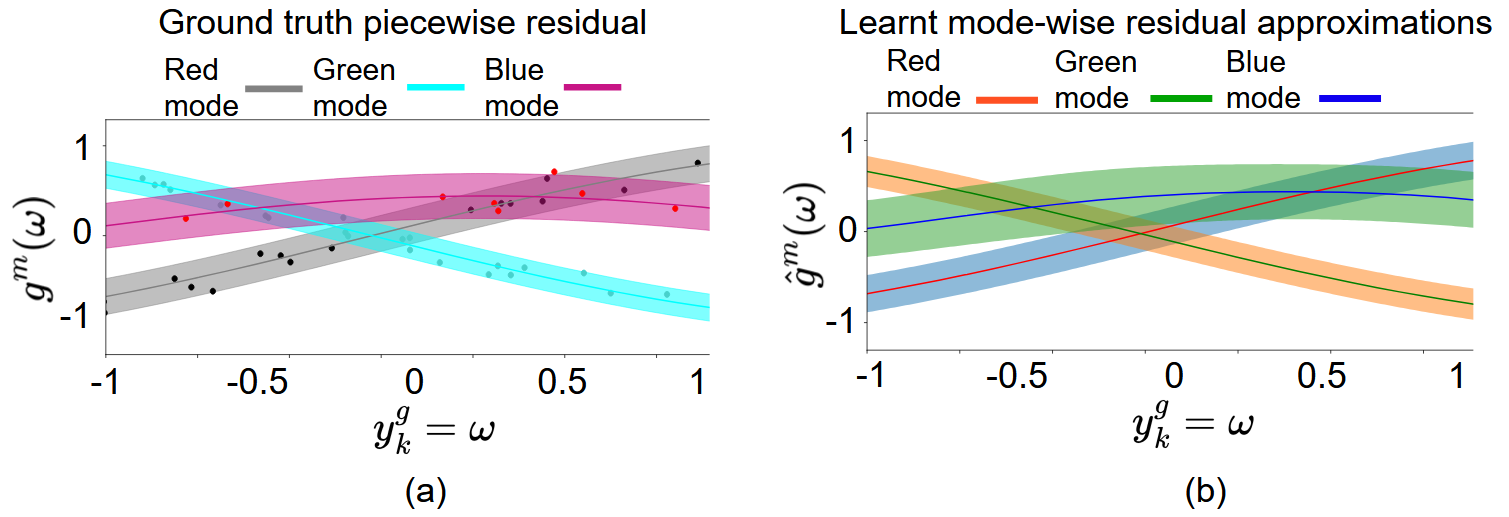}}
\end{tabular}
\vspace{-10pt}
\caption{\small{(a) Ground truth and (b) learnt \gls{gp} mode residuals for each mode present in the workspace. The mode names are assigned based on the mode colors depicted in Fig~\ref{fig:mode_mapping_final}.}}
\label{fig:true_func_quad2d}
\vspace{-20pt}
\end{figure}

We demonstrate the proposed algorithm on a 2-D quadrotor system~\cite{safe_control_gym} with $X\subset\mathbb{R}^6, U\subset\mathbb{R}^2$. 
Here, $\ydk = (x^1_k, x^2_k)$ corresponding to the position of the robot in the $x-z$ plane and $\ygk=\omega$ (angular velocity). The residual is as depicted in Figure~\ref{fig:true_func_quad2d} and affects only the $x$ position dynamics \ie $\Bg = [1, 0, 0, 0, 0, 0]^T$ to yield a disturbed kinematic model similar to that considered in~\cite{dist_kin_model}. The cost matrix $Q \in \mathbb{R}^{6\times6}$ (and p.s.d) is chosen to prioritize $x-z$ tracking with $R = 0.01I_{2\times2}$ for the inputs. 

\begin{figure}[h!]
\centering
\begin{tabular}{cccc}
{\includegraphics[width=0.45\textwidth, trim={0cm 0cm 0cm 0cm},clip]{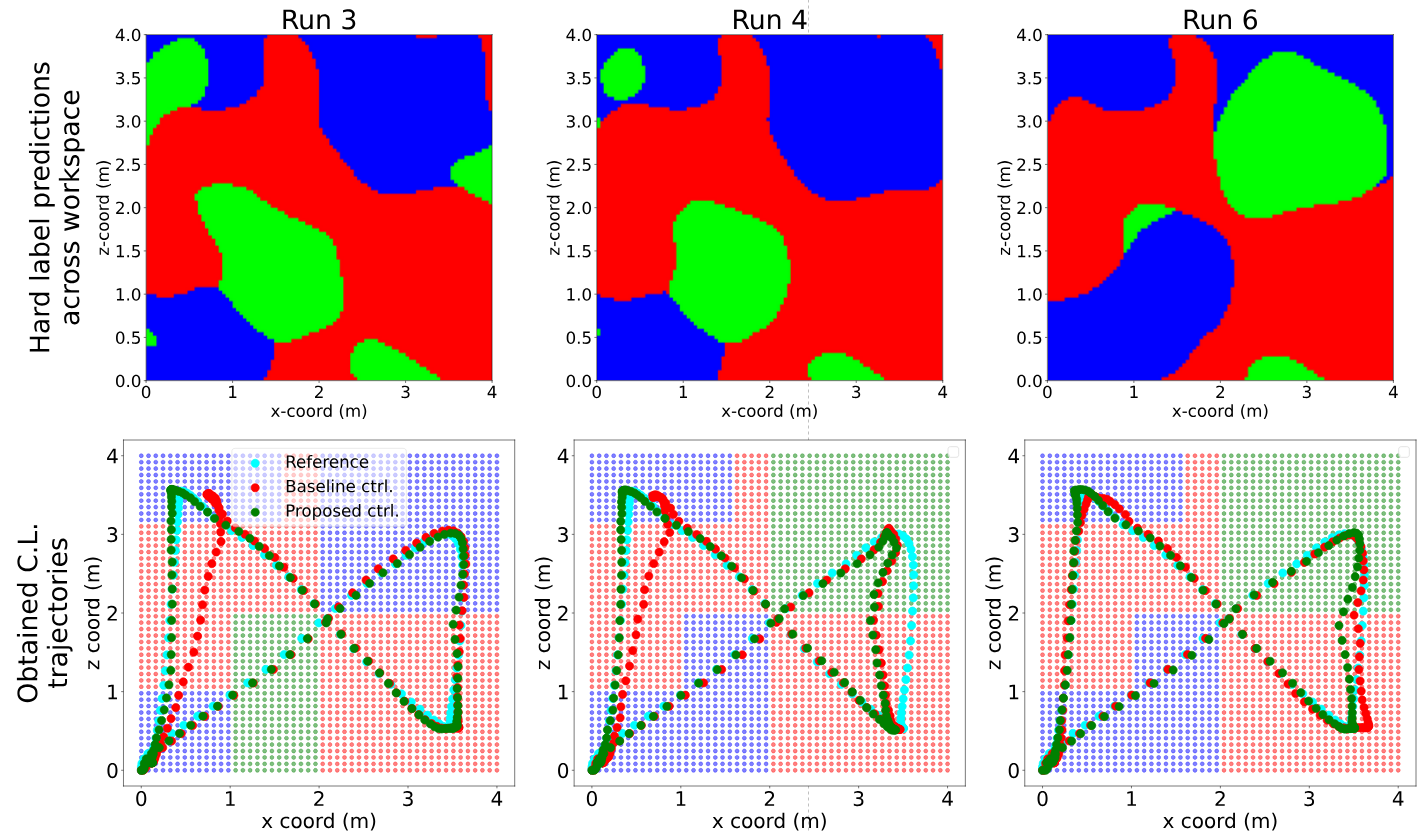}}
\end{tabular} \vspace{-7pt}
\caption{\small{(Top) Plots of one-hot vectors $\deltahatrk$~\eqref{eq:deltahat_ref_gen} (for $\ydk$ uniformly sampled across the $x-z$ bounds) predicted by $\hat{\delta}$, at the \emph{current} run. These plots are shown for the baseline controller's run to demonstrate that it can learn the right distribution but does not perform as well as the proposed solution. (Bottom) Plots of the $x-z$ trajectory for the quadrotor system across repeated runs of a Figure-8 tracking task plotted on ground-truth mode distribution at that run. Note the mode distribution shift between Runs 3 and 4.}}
\label{fig:mode_mapping_final}
\end{figure}

For the experiment, we first run an \gls{mpc} controller, using \emph{only} the nominal model, on an initial trajectory due to lack of prior mode distribution information. Henceforth, we switch to Figure-8 tracking task (as seen in Fig~\ref{fig:mode_mapping_final}) and execute repeated runs of the same. \emph{After} run 3, the active mode changes in some parts of the $x-z$ workspace. We discuss how our approach handles for this using both control cost and model prediction accuracy metrics.

\noindent\textbf{Discussion: } The results of the experiment are visualized in Fig~\ref{fig:mode_mapping_final}. The average \gls{ol} computation time for the proposed controller runs was $\sim$99 ms, comparable to that reported in Table~\ref{table:int_tracking_lti}, showing that the tractability scales with the dimension of the state.

\begin{figure}[h!]
\centering
\begin{tabular}{cccc}
{\includegraphics[width=0.53\textwidth, trim={2cm 0.2cm 0cm 0cm},clip]{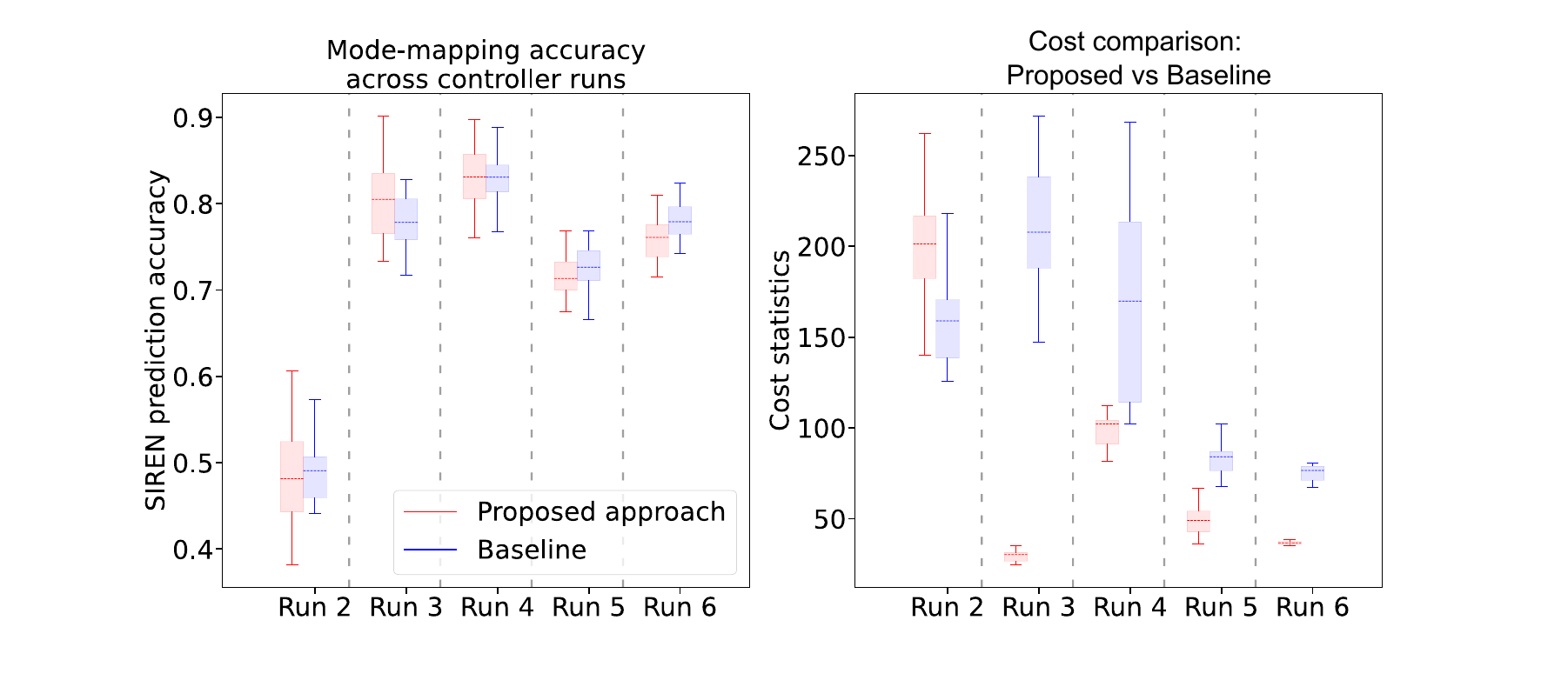}}
\end{tabular} \vspace{-7pt}
\caption{\small{(Left) $\hat{\delta}$ prediction accuracy \emph{by argmax}~\eqref{eq:deltahat_ref_gen} on 10K uniform samples across the workspace. (Right) Cost statistics over successive iterations of the repetitive task (over 30 simulation runs). The dotted red and blue lines are the mean \gls{cl} cost for the proposed and baseline controllers respectively.}}
\label{fig:mapping_results_boxplot}
\end{figure}

The statistics of the experiment over 30 runs of each controller are summarized in Fig~\ref{fig:mapping_results_boxplot}. Run 1 uses a nominal controller for initial sample collection and is not included in Fig~\ref{fig:mapping_results_boxplot}. The control performance is also affected when the prediction model accuracy is low over the reference trajectory, as seen in Run 2. The mode transition occurs at the end of Run 3, translating to an increase in control cost for Run 4 due to the newly inaccurate prediction model. After re-training on new samples using the proposed trade-off update rule~\eqref{eq:tradeoff}, the control cost falls again due to the improved prediction accuracy of $\hat{\delta}$. Though both the baseline and the proposed controller runs demonstrate similar $\hat{\delta}$ accuracy, the proposed controller is able to obtain lower average control cost because of allowed mode switching over the \gls{ol} optimization horizon. We also note that while the prediction accuracy of $\hat{\delta}$ stagnates towards the final run (as seen in Fig~\ref{fig:mapping_results_boxplot}), the accuracy in a region \emph{around the repetitive task} is sufficiently high to have generated iterative improvements in control performance.

\begin{figure}[h!]
\centering
\begin{tabular}{cccc}
{\includegraphics[width=0.45\textwidth, trim={0cm 0cm 0cm 0cm},clip]{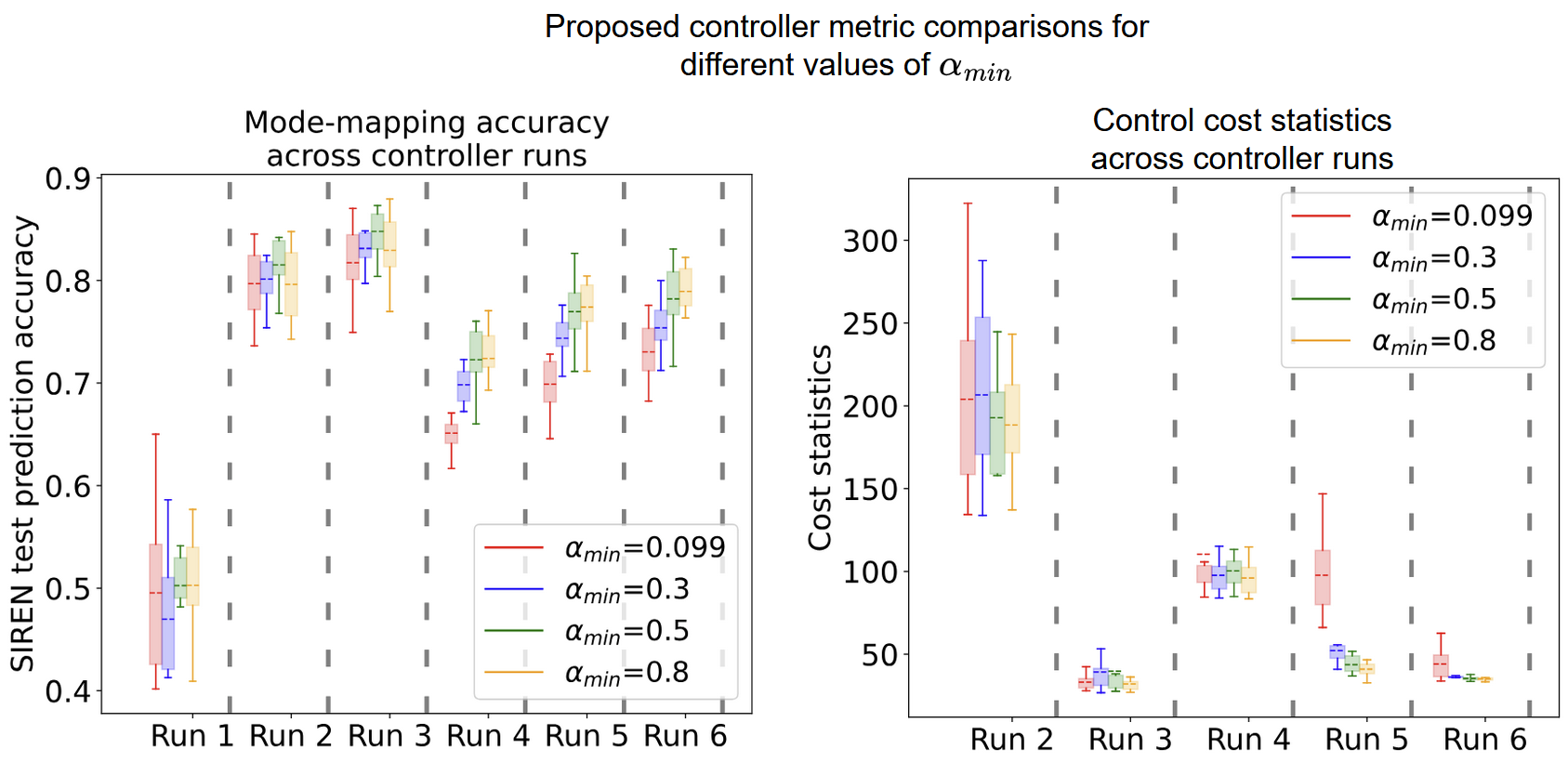}}
\end{tabular} \vspace{-7pt}
\caption{\small{Comparison of hard label prediction accuracy and closed-loop cost of the proposed control approach over 20 runs for different values of $\alphamin$.}}
\label{fig:alphamin_ablation_demo}
\end{figure}

Finally, Figure~\ref{fig:alphamin_ablation_demo} demonstrates the effect of $\alphamin$ on the speed at which the predictor $\hat{\delta}$ updates to changes in the mode distribution. It can be seen that after the mode change at Run 4, the prediction accuracy statistics for $\alphamin \in \{0.099, 0.3\}$ trail behind those for $\alphamin \in \{0.5, 0.8\}$. This signifies that lower $\alphamin$ translates to slower adaptation. $\alphamin=0.99$ resulted in higher closed-loop cost in the runs after the mode shift (particularly noticeable in run 5).

\section{Conclusions and Future Work}
We presented a computationally tractable method for controlling a system subject to \gls{pwr} dynamics in an apriori unseen environment with a \emph{time-varying} mode distribution. Through simulations, we demonstrate that the heuristic approximations proposed to reduce an \gls{minlp} optimization problem to an \gls{nlp} show improvements in computational tractability while maintaining reasonable control performance. We show that the approach is capable of effectively trading off prior prediction estimates (learnt using previous trajectory data samples) with likelihood estimates obtained from the latest run to adapt to distribution changes over time. 

Despite the improvements in performance we demonstrate, the method has limitations. As highlighted by Assumption~\ref{as:apriori_gp}, the proposed approach relies on knowing a residual approximation model for all modes we might encounter, which can be limiting in practice. Also, it is crucial that data collection for mode identification in unseen environments is done with \emph{safety}. Ongoing work involves drawing on the literature concerning hybrid mode identification to relax Assumption~\ref{as:apriori_gp} while also improving safety during the data collection process.
\section{Acknowledgements}
The authors would like to acknowledge Shamak Dutta for his help with proofreading the paper.

\bibliographystyle{IEEEtran}
\bibliography{bibs/acc2025}
\end{document}